\documentclass[12pt]{iopart}
\usepackage{iopams}
\usepackage[square,comma,sort&compress,numbers]{natbib}
\usepackage[colorlinks,linkcolor=blue,anchorcolor=blue,citecolor=blue,filecolor=blue,menucolor=blue,urlcolor=blue]{hyperref}

\newcommand{\hr}{{\cal H}}
\newcommand{\gset}{{\cal G}}
\newcommand{\da}{\Delta_{\! A}}

\newcommand{\dda}{\delta \! A}

\newcommand{\eff}{\mathrm{eff}}

\providecommand{\norm}[1]{\|#1\|}

\begin{document}

\title{Equilibration of isolated macroscopic quantum systems}
\author{Peter Reimann$^{1}$ and Michael Kastner$^{2}$}
\address{$^{1}$Universit\"at Bielefeld, Fakult\"at f\"ur Physik, 33615 Bielefeld, Germany}
\address{$^{2}$ National Institute for Theoretical Physics (NITheP), 
Stellenbosch 7600, South Africa and
Institute of Theoretical Physics, University of Stellenbosch, 
Stellenbosch 7600, South Africa\eads{\mailto{reimann@physik.uni-bielefeld.de}, \mailto{kastner@sun.ac.za}}}

\begin{abstract}
We investigate the equilibration of an isolated 
macroscopic quantum system in the sense 
that deviations from a
steady state become unmeasurably 
small for the overwhelming majority 
of times within any sufficiently large
time interval.
The main requirements are that the initial 
state, possibly far from equilibrium, 
exhibits a macroscopic population 
of at most one energy level and that
degeneracies of energy eigenvalues and 
of energy gaps (differences of energy eigenvalues)
are not of exceedingly large 
multiplicities.
Our approach closely follows and extends 
recent works by Short and Farrelly [2012 New J.\ Phys.\ {\bf 14} 013063], 
in particular going beyond the realm
of finite-dimensional systems and
large effective dimensions.
\end{abstract}


\maketitle
\tableofcontents

\section{Introduction}
\label{s1}
It is a basic everyday experience that 
isolated macroscopic systems, i.e.\ systems 
consisting of many particles or other 
microscopic degrees of freedom, approach some steady
equilibrium state after a sufficiently long
time evolution, no matter how far from 
equilibrium they started out.
More precisely, for every single run of
the experiment, one still may encounter 
certain statistical or quantum mechanical
fluctuations, especially
for microscopic observables, but on the 
average over many repetitions of the 
experiment, all expectation values 
appear to equilibrate.

The reconciliation of this irreversible
behavior with quantum mechanical re\-vers\-ibil\-i\-ty 
and revival/recurrence properties is a
long standing problem \cite{skl93}
which has recently be reconsidered 
from a new viewpoint,
without focusing on any specific model class
and without any modification/approximation of the
exact quantum mechanical time evolution 
\cite{rei08,lin09,lin10,rei10,sho11,sho12}.
The key point of these works is to show that
the expectation values may still exhibit 
everlasting small fluctuations around their equilibrium values, as well as very rare
large excursions away from equilibrium (including the above-mentioned
recurrences), but quantitatively these 
fluctuations are either unobservably small
compared to any reasonably achievable 
resolution limit, or exceedingly rare
on any realistic time scale
after initial transients have died out.
In this sense, the system indeed equilibrates.

Originally, these conclusions have been based 
on the following main assumptions \cite{rei08,lin09,lin10,rei10}:
\begin{enumerate}
\item The considered observables represent
experimental measurement devices with a 
finite instrumental range and a possibly 
very small but still `reasonable' 
resolution limit.
\item The initial condition exhibits a
small occupation probability of every 
single energy level, which is very plausible 
in view of the unimaginably large level 
density of typical macroscopic systems.
\item All energy eigenvalues are non-degenerate.
\item All energy gaps are non-degenerate, 
i.e.\ a pair of distinct energy eigenvalues 
never exhibits the same energy difference as 
some other pair.
\item Either the considered Hilbert space 
must be finite-dimensional 
\cite{lin09,lin10},
or some formal manipulations are
not rigorously justified \cite{rei08,rei10},
in particular interchanging the limit of 
infinite dimensions with the long-time limit.
\end{enumerate}

The restrictions (iii) and (iv) have recently 
been overcome in two very important
contributions by Short \cite{sho11}
and by Short and Farrelly \cite{sho12}.
In our present work, we closely follow
and further extend their approach
by relaxing also the above conditions (ii) and (v).
Namely, we only require that the second largest
level population must be small, while
the occupation probability of one level
may be macroscopic (non-small).
Such a case may e.g.\ be of relevance
for thermal equilibrium states at
extremely low temperatures.
Furthermore, we will admit and treat with 
care (countably) infinite
dimensional systems.

Our present approach also bears resemblance
to recent advancements made by Goldstein 
and coworkers of ideas originally due to von Neumann
\cite{gol10a,gol10b,gol10c,tas10}.
The main difference is that in these works
an alternative notion is adopted of when a system 
is in or close to equilibrium,
and the main emphasis is laid on macroscopic
(coarse-grained) observables, exhibiting
the same expectation value for most
states within any quantum mechanical 
energy shell.

Furthermore, our approach is complementary 
to numerous recent investigations
of equilibration for various specific systems, 
observables, and initial conditions, 
and often with a main focus on the role of 
(non-)integrability, see e.g.\ 
\cite{tas98,sre99,caz06,rig07,col07,man07,cra08,bar08,rig08,spec,pal10,ban11,pol11,camp11,jacobson11,kas11,kas12}
and references therein.

Another issue closely related to 
equilibration is the problem of 
thermalization, i.e.,
the question whether, and to what extend, 
the above mentioned equilibrium states
agree with any one of Gibbs's statistical
ensembles.
This important issue, either for an
isolated system {\em per se} or
for an isolated system-plus-bath 
composite, has been recently 
addressed in e.g., 
\cite{rei08,lin09,lin10,rei10,gol10a,gol10b,gol10c,tas10,tas98,sre99,caz06,rig07,col07,man07,cra08,bar08,rig08,spec,pal10,gog11,ban11,pol11,ike11},
but will not be considered here in any 
further detail.
In other words, our notion 
of {\em equilibrium}\/ is weaker than that of
{\em thermal equilibrium}\/.

\section{General Framework}
\label{s2}
\subsection{System and Hamiltonian}
\label{s21}
We consider an isolated system,
confined to a finite region of space 
and involving a finite 
number of particles.
Later we will mainly be 
interested in macroscopic
systems, but for the moment
any finite number of degrees of 
freedom is admitted.
In particular, we may  
be dealing with a compound system, 
consisting of a subsystem of actual 
interest and its environment 
(reservoir or thermal bath).

According to standard quantum mechanics, 
such a system is modeled by a
time-independent Hamiltonian $H$ 
on a separable (i.e.\ at most countably
infinite-dimensional) Hilbert space $\hr$. 
Since we consider the system to
be confined to a finite region of space, 
all eigenvectors of
$H$ represent bound states and the 
spectrum of $H$ is discrete (pure point).
As a consequence, the Hamiltonian can be written in the form
\begin{equation}
H = \sum_n E_n P_n,
\label{1}
\end{equation}
where the $P_n$ are projectors onto the 
eigenspaces of $H$ with eigenvalues $E_n$, 
satisfying
\begin{eqnarray}
P_m P_n = \delta_{mn} P_n,
\label{2}
\\
\sum_n P_n = 1,
\label{3}
\\
E_n \neq E_m \quad \mbox{if $n\neq m$}.
\label{4}
\end{eqnarray}
Here, $n,m\in\{1,\dots,d_E\}$, where the 
number $d_E$ of distinct energy eigenvalues may be
finite or infinite. The symbol $\sum_n$
indicates a summation over all those $n$-values, 
$\delta_{mn}$ is the Kronecker symbol, 
and $1$ the identity on $\hr$.
In particular, any energy eigenvalue $E_n$ is allowed to
be degenerate and its multiplicity is given by
\begin{equation}
\mu_n:=\Tr \{ P_n \},
\label{5}
\end{equation}
where $\Tr$ denotes the trace on $\hr$.
The dimension of $\hr$ thus amounts to 
$\sum_n \mu_n$ and may be finite or infinite.

\subsection{States and dynamics}
\label{s22}
The system's state at time $t$ is captured as usual
by a density operator $\rho (t)$, describing 
either a statistical ensemble (mixed state)
or a pure state, and evolving in time 
according to 
$\rho(t)=U_t\rho(0) U_t^\dagger$
with time-evolution operator $U_t:=\exp\{-\rmi Ht\}$ and $\hbar =1$.
With \eref{1} we can conclude that
\begin{equation}
\rho(t)=\sum_{mn} \rho_{mn}(0) \exp[-\rmi (E_m-E_n)t],
\label{6}
\end{equation}
where we have introduced the auxiliary operators
\begin{equation}
\rho_{mn}(t):=P_m\rho(t)P_n.
\label{7}
\end{equation}
While $\rho_{nn}(t)$ are thus time-independent and
self-adjoint operators, the same generically 
does not hold for $\rho_{mn}(t)$ when $m\neq n$.
In particular, $\rho_{nn}(t)$ equals $\rho_{nn}(0)$
and the time-argument will often be omitted.

\subsection{Realistic observables}
\label{s23}
Observables are represented as usual by self-adjoint
operators $A$ with expectation values
$\Tr\{\rho(t)A\}$.
In order to model real experimental measurements 
it is however
not necessary to admit any arbitrary self-adjoint 
operator \cite{realobs1,realobs2,realobs3,realobs4,realobs5,lof,geo95,pop06}.
Rather, it is sufficient to focus on
{\em realistic observables} 
in the following sense \cite{rei08,rei08a,rei10}: 
Any observable $A$ must represent an experimental 
device with a {\em finite range} of possible 
outcomes of a measurement,
\begin{equation}
\da := 
\sup_{\psi\in S(\hr)} \langle\psi|A|\psi\rangle
- \inf_{\psi\in S(\hr)} \langle\psi|A|\psi\rangle
= a_{\sup} - a_{\inf} < \infty,
\label{8}
\end{equation}
where
\begin{equation}
S(\hr):=\left\{\psi\in\hr\,\big|\,\langle\psi|\psi\rangle=1\right\}\subset \hr
\end{equation}
denotes the set of normalized vectors in $\hr$.
Moreover, this working range $\da$
of the device must be limited to experimentally 
reasonable values compared to its 
{\em resolution limit $\dda$}.
All measurements known to the present authors 
yield less than 20 relevant digits, i.e.\ 
\begin{equation}
\da/\dda \leqslant 10^{20}.
\label{9}
\end{equation}
Maybe some day 100 or 1000 relevant digits will
become feasible, but it seems reasonable that a theory
which does not go very much beyond that will do.
Note that similar restrictions also apply
to `numerical experiments' by 
computer simulations.

According to \eref{8}, all eigenvalues of $A$
must be contained within the finite interval 
$[a_{\inf}, a_{\sup}]$ and the operator norm
\begin{equation}
\norm{A}:=\sup_{\psi\in S(\hr)} \norm{A|\psi\rangle}
\label{10}
\end{equation}
is finite and equal to
$\max\{|a_{\inf}|, |a_{\sup}|\}$.
(As usual, the vector norm on the right hand side 
of \eref{10} is the one induced by 
the scalar product on $\hr$).

\subsection{Level populations}
\label{s24}
The specific observable $A=P_n$ describes
the population of the energy level $E_n$ with
expectation value (occupation probability)
\begin{equation}
p_n:=\Tr\{ P_n \rho(t) \}=\Tr \{ \rho_{nn} \}.
\label{11}
\end{equation}
The last relation shows the time-independence 
of $p_n$ and follows from \eref{2}, \eref{7},
and the invariance of the trace under cyclic 
permutations.

For a system with $f$ degrees of freedom
there are roughly $10^{\Or(f)}$ energy eigenstates 
with eigenvalues in every interval of $1$J beyond the ground 
state energy (\cite{lldiu}; for a more detailed discussion,
see also section 2.1 of \cite{rei10}).
The same estimate carries over to the number
of energy eigenvalues under the assumption that
their multiplicities \eref{5}
are much smaller than $10^{\Or(f)}$.
For a macroscopic system with
$f=\Or(10^{23})$, the energy levels are thus 
unimaginably dense on any decent energy 
scale and even the most careful experimentalist 
will not be able to populate only a few of them
with significant probabilities $p_n$.

We recall that $\rho(t)$ may be a pure state,
but the case of foremost interest is on
mixed states describing a statistical 
ensemble.
Then, $p_n$ describes an ensemble 
average over many repetitions of the experiment.
Hence, an `accidentally large' 
population of a few levels in 
one particular experimental run is still admissible,
but very unlikely to occur again when the experiment is repeated.

To obtain a rough estimate, we 
imagine that there are exactly 
$10^{(10^{23})}$ energy levels per J.
Even if the experimentalist can prepare the 
energy of the system with a fantastically small
uncertainty of $10^{-(10^{22})}$J, there still 
remain $N:=10^{0.9\times 10^{23}}$ energy levels 
which may be occupied with significant 
probabilities.
If all of them are populated equally,
we obtain $p_n=1/N$
for $N$ of the indices $n$, and 
$p_n=0$ for all other $n$.
If not all $N$ levels are populated
equally, but rather any $p_n$ 
may assume arbitrary 
values between zero and $10^{(10^{22})}$
times the average population 
$1/N$, we still obtain
$p_n\leqslant 10^{-0.8\times 10^{23}}$.
Returning to the general case, we can conclude 
\cite{rei08,rei08a,rei10}
that even if the system's energy is fixed up to
an extremely small experimental uncertainty,
and even if the energy levels are populated 
extremely unequally, we still expect that 
even the largest ensemble-averaged 
level population $p_n$ will be extremely 
small and typically satisfy the 
rough estimate
\begin{equation}
\max_n p_n=10^{-\Or(f)}.
\label{12}
\end{equation}

\subsection{Macroscopic population of one energy level}
\label{s25}
There is one physically significant situation
in which the above arguments 
may become questionable.
Namely, for an isolated macroscopic
system which 
approaches a thermal equilibrium state with
an extremely low temperature, 
it might be conceivable that the ground state 
energy exhibits a macroscopic population, 
i.e.\ the corresponding $p_n$ is no longer 
extremely small.
Hence, we should omit that specific
$p_n$ in the maximization \eref{12},
formally written as
\begin{equation}
{\max_n}' p_n=10^{-\Or(f)}.
\label{13}
\end{equation}
In other words, the prime indicates that
the largest $p_n$ is not included into the
maximization and hence
${\max_n}' p_n$ {\em represents the 
second largest level population}.

Further situations resulting in a non-small population 
of one single level may be caused e.g.\ by certain
`gaps' in the energy spectrum or by one
level with an extremely high multiplicity \eref{5}.

Note that the expected relations 
$p_n\geqslant 0$ and $\sum_n p_n=1$ readily follow 
from \eref{3}, \eref{11}, and the fact 
that $\rho(t)$ is non-negative and of unit trace.
We thus can conclude that the maxima $\max_n p_n$ and 
${\max_n'} p_n$ indeed exists---as anticipated in 
\eref{12} and \eref{13}---and that 
$0\leqslant {\sum_n'} p_n<1$, where the prime 
in ${\sum_n'}$ excludes the index 
$n$ belonging to the maximal $p_n$.
It follows that
\begin{eqnarray}
{\sum_n}' p_n^2 \leqslant {\max_n}' p_n
{\sum_n}' p_n \leqslant {\max_n}' p_n,
\label{14}
\\
{\max_n}' p_n = \big( {\max_n}' p_n^2 \big)^{1/2} 
\leqslant \big( {\sum_n}' p_n^2 \big)^{1/2},
\label{15}
\end{eqnarray}
and we can conclude that
\begin{equation}
{\max_n}' p_n \; \mbox{small}\ \Leftrightarrow\ 
{\sum_n}' p_n^2\; \mbox{small}.
\label{16}
\end{equation}

In references \cite{lin09,lin10,sho11,sho12}, the quantity
$d_{\eff}:=1/\sum_n p_n^2$, called the effective dimension
of the state $\rho(t)$, is introduced. It quantifies the number of distinct 
energies that contribute notably to this state,
and is required to be a large number.
Observing that the equivalence 
\eref{16} also applies without primes, we see that 
the requirement of a large effective dimension is
fulfilled if and only if the maximal level 
population $\max_n p_n$ is small.
However, in the more general case including the 
primes in \eref{16}, as considered in our present 
work, the effective dimension $d_{\eff}$ may not 
be large any more.

\subsection{Equilibration and equilibrium ensemble}
\label{s26}
Generically, the statistical ensemble 
$\rho(t)$ is not stationary right from 
the beginning, in particular for an initial
condition $\rho(0)$ out of equilibrium.
But if the right hand side of \eref{6}
depends on $t$ initially, it cannot 
approach for large $t$ any time-independent 
`equilibrium ensemble' whatsoever.
In fact, any mixed state $\rho(t)$ 
returns arbitrarily close (with respect to some 
suitable distance measure in Hilbert space) 
to its initial state $\rho(0)$ for certain, 
sufficiently large times $t$, 
as demonstrated for instance in 
appendix D of \cite{hob71}.

We will therefore focus on the weaker 
notion of equilibration outlined in
\sref{s1}, requiring the
existence of a time-independent 
`equilibrium state' $\omega$ (density operator) 
with the property that the difference
\begin{equation}
\sigma(t):=\Tr\{\rho(t) A\}-\Tr\{\omega A\}
\label{17}
\end{equation}
between the true expectation value 
$\Tr\{\rho(t) A\}$ and the equilibrium reference 
value $\Tr\{\omega A\}$ is unresolvably 
small for the overwhelming majority 
of times $t$ contained in any sufficiently
large (but finite) time interval $[0,T]$.
(Note that initial transients become irrelevant 
if $T$ is chosen large enough.)

Heuristically, if any such equilibrium ensemble
exists, then it should be given by the
infinite time average of $\rho(t)$.
In view of \eref{6} this suggests the 
definition
\begin{equation}
\omega:=\sum_n\rho_{nn}.
\label{18}
\end{equation}
However, from a more rigorous viewpoint, it is not
so obvious that averaging \eref{6} over arbitrary 
but finite times leads to a well-defined long-time
limit, which is furthermore given by \eref{18}.
Specifically, for infinite-dimensional systems,
interchanging the infinite time limit with the
infinite double-sum in \eref{6} is problematic.
We avoid all these difficulties by defining 
$\omega$ according to \eref{18} 
without any reference to averages 
over time. An alternative and entirely unproblematic viewpoint
is to consider $\omega$ as the time-independent
part of $\rho(t)$.

One readily sees that $\omega$ inherits from $\rho(t)$
the properties of being self-adjoint, non-negative, 
and of unit trace.
Furthermore, $\omega$ satisfies the trivial 
time evolution $U_t \omega U_t^\dagger=\omega$.
In other words, $\omega$ is indeed a perfectly 
well-defined density operator.

We finally note that, as far as the differences in \eref{17}
are concerned, nothing changes if $A$ is replaced by $A+c\, 1$ 
with an arbitrary real $c$. 
Thus, we henceforth can assume without loss of 
generality that $a_{\sup}=-a_{\inf}$ in
\eref{8}, implying with \eref{10} that
\begin{equation}
\norm{A}=\da /2.
\label{19}
\end{equation}

\section{From infinite to finite dimensions}
\label{s3}
We focus on infinite-dimensional Hilbert spaces $\hr$
and denote the normalized eigenvectors of the Hamiltonian $H$
by $|\nu\rangle$ with $\nu=1,2,\dots$.
For any given positive integer $d$ we define the projectors
\begin{eqnarray}
P := \sum_{\nu =1}^d |\nu\rangle\langle\nu|,
\label{20}
\\
Q := 1-P.
\label{21}
\end{eqnarray}
For an arbitrary density operator $\rho$ and any
observable $A$ it follows that
\begin{equation}
\Tr\{\rho  A\} = \Tr\{(P+Q)\rho (P+Q) A\}= R_1+R_2+R_3
\label{22}
\end{equation}
with
\begin{eqnarray}
R_1 := \Tr\{ P \rho P A \},
\label{23}
\\
R_2 := \Tr\{Q \rho A\},
\label{24}
\\
R_3 := \Tr\{P \rho Q A\}.
\label{25}
\end{eqnarray}
Making use of $P^2=P$ and the cyclic invariance 
of the trace, one can rewrite $R_1$ as 
$\Tr\{(P \rho P) (P A P)\}$
or
\begin{eqnarray}
R_1 & \phantom{:}= \Tr\{\tilde\rho\tilde A\},
\label{26}
\\
\tilde\rho & := P \rho P,
\label{27}
\\
\tilde A & := P A P.
\label{28}
\end{eqnarray}

To further evaluate $R_2$ we represent the self-adjoint 
operator $\rho$ in terms of its eigenvalues $\rho_\nu$ and 
eigenvectors $|\phi_\nu\rangle$,
\begin{equation}
\rho = \sum_{\nu=1}^\infty \rho_\nu|\phi_\nu\rangle\langle\phi_\nu|. 
\label{29}
\end{equation}
Since $\rho$ is non-negative, all $\rho_\nu$ are non-negative and
\begin{equation}
\rho^{1/2} := 
\sum_{\nu=1}^\infty \sqrt{\rho_\nu}\ 
|\phi_\nu\rangle\langle\phi_\nu|
\label{30}
\end{equation}
is a well-defined, self-adjoint operator with the property
that $\rho^{1/2} \rho^{1/2}= \rho$.
It follows that
\begin{equation}
|R_2|^2 = \bigl|\Tr\{(Q \rho^{1/2}) (\rho^{1/2} A)\}\bigr|^2
\leqslant \Tr\{Q\rho Q\} \Tr\{A \rho A\},
\label{31}
\end{equation}
where we exploited the Cauchy--Schwarz inequality 
\begin{equation}
\bigl|\Tr\{B^\dagger C\}\bigr|^2\leqslant \Tr\{B^\dagger B \}  \Tr\{C^\dagger C \}
\label{32}
\end{equation}
for the scalar product $\Tr\{B^\dagger C\}$ 
of arbitrary operators $B$ and $C$ 
(for which all traces in \eref{32} exist).
The last term in \eref{31} equals $\Tr\{\rho A^2\}$, and by evaluating
the trace by means of the orthonormal basis $|\phi_n\rangle$ one can
infer with \eref{10} and \eref{29} that
\begin{equation}
\Tr\{\rho A^2\}\leqslant \norm{A^2}  \Tr\{\rho\}\leqslant \norm{A}^2,
\label{33}
\end{equation}
where we exploited that $\Tr\{\rho\}=1$ and $\norm{A^2}=\norm{A}^2$
in the last relation.
Finally, we conclude from \eref{20} and \eref{21} that
\begin{equation}
\Tr\{Q\rho Q\} = \sum_{\nu=d+1}^\infty \langle\nu|\rho|\nu\rangle
\label{34}
\end{equation}
and hence
\begin{equation}
|R_2|^2 \leqslant\norm{A}^2 \sum_{\nu=d+1}^\infty \langle\nu|\rho|\nu\rangle.
\end{equation}

Next, starting from \eref{25}, we rewrite $R_3$ as $\Tr\{AP\rho Q\}$.
Noting that all four operators under this trace are self-adjoint
and that $\Tr\{B^\dagger\}=\Tr\{B\}^\ast$ for arbitrary 
operators $B$, we can conclude that $|R_3|=|\Tr\{Q\rho PA\}|$.
Proceeding similarly as in \eref{31}--\eref{33} and using $\Tr\{\tilde{\rho}\}\leqslant1$ one 
finds that
\begin{equation}
|R_3|^2\leqslant\Tr\{Q\rho Q\}\norm{A}^2 \Tr\{\tilde{\rho}\}\leqslant \norm{A}^2 \sum_{\nu=d+1}^\infty \langle\nu|\rho|\nu\rangle.
\end{equation}
This in turn implies via \eref{22}--\eref{28} that
\begin{equation}\label{35}
|\Tr\{\rho A\}-\Tr\{\tilde\rho\tilde A\}|\leqslant 2 \norm{A} \Bigl(\sum_{\nu=d+1}^\infty \langle\nu|\rho|\nu\rangle\Bigr)^{1/2}
\end{equation}
for arbitrary density operators $\rho$.
Applying this relation to the specific density operators
$\rho(t)$ in \eref{6} and $\omega$ in \eref{18}
we obtain
\begin{equation}
|\Tr\{\rho(t) A\}-\Tr\{\omega A\}| =
|\Tr\{\tilde\rho(t) \tilde A\}-\Tr\{\tilde \omega \tilde A\}| + R
\label{36}
\end{equation}
with
\begin{eqnarray}
R &\leqslant |\Tr\{\rho(t) A\}-\Tr\{\omega A\} 
- \Tr\{\tilde\rho(t) \tilde A\}+\Tr\{\tilde \omega \tilde A\}|
\label{37}\\
& \leqslant 4\norm{A} \Big(\sum_{\nu=d+1}^\infty 
\langle\nu|\rho (0) |\nu\rangle\Big)^{1/2}.
\label{38}
\end{eqnarray}
In \eref{37} we exploited the fact that $|x|-|y|\leqslant |x-y|$ 
for arbitrary real $x$ and $y$, while 
in \eref{38} we applied
\eref{35} to $\rho(t)$ and $\omega$ and we 
took into account that, according to
\eref{6} and \eref{18}, both
$\langle \nu | \rho(t)|\nu\rangle$ and
$\langle \nu | \omega|\nu\rangle$ are equal to
$\langle \nu |\rho(0)|\nu\rangle$.

Observing that $\sum_{\nu=1}^d\langle \nu |\rho(0)|\nu\rangle$
increases with $d$ and approaches $\Tr\{\rho(0)\}=1$
for $d\to\infty$ it follows that, for any given $\epsilon>0$,
there exists a finite $d(\epsilon)$ with the property that
$R\leqslant \norm{A}\epsilon$. 
According to \eref{9}, upon choosing $\epsilon=10^{-20}$
and observing \eref{19}, we can conclude that
\begin{equation}
R\leqslant \dda/2
\label{39}
\end{equation}
with one common, finite $d$ for all 
experimentally realistic observables $A$.

As far as experimentally resolvable differences
\eref{17} are concerned, it thus
fol\-lows from \eref{36} and \eref{39} that
it is sufficient to consider, instead of
$\rho(t)$ and $A$, their counterparts
$\tilde \rho(t)$ and $\tilde A$.
According to \eref{19}, \eref{27} and \eref{28}
these are the projections/restrictions of the original
operators $\rho(t)$ and $A$ onto the finite
dimensional sub-Hilbert space $\tilde\hr\subset \hr$, 
spanned by the first $d$ energy eigenvectors
$\{|\nu\rangle\}_{\nu=1}^d$.
Note that while $\tilde\hr$ is independent of $A$, 
it does depend on $\rho(0)$ and $H$ in general.

It remains to be shown that the entire framework set out in
\sref{s2} can be consistently restricted to
the finite-dimensional Hilbert space $\tilde \hr$:
Observing that the projectors $P_n$ from
\eref{1} commute with the projector $P$ 
defined in \eref{20} implies that
\begin{equation}
\tilde P_n:=P P_n P=P P_n=P_n P.
\label{40}
\end{equation}
Setting $\tilde H:=PHP$ and keeping only indices $n$
with non-zero $\tilde P_n$, relations
\eref{1}--\eref{5} remain valid, but now with
finite sums $\sum_n$ and finite multiplicities
$\tilde \mu_n$.
Furthermore, one sees that $\tilde H$ indeed reproduces the
correct time evolution of $\tilde \rho(t)$ with finite 
sums in \eref{6}.
While $\tilde \rho(t)$ is still non-negative 
and self-adjoint, the trace now satisfies
\begin{equation}
\Tr\{\tilde \rho(t)\}=\sum_n\tilde p_n\leqslant 1.
\label{41}
\end{equation}
Likewise, with respect to the 
operator norm \eref{10} and the level population \eref{11} 
one finds that
\begin{equation}
\norm{\tilde A} \leqslant \norm{A},\qquad
\tilde p_n \leqslant p_n. 
\label{43}
\end{equation}

\section{Finite-dimensional systems}
\label{s4}
The main objective of this section is to establish bounds
on the difference
\begin{equation}
\tilde \sigma(t) :=\Tr\{\tilde \rho(t)\tilde A\}
-\Tr\{\tilde\omega \tilde A\},
\label{44}
\end{equation}
where tildes indicate the projections/restrictions to the 
finite-dimensional Hilbert space $\tilde \hr$ from the
previous subsection in case the 
original Hilbert space $\hr $ 
was infinite-dimensional (otherwise the tildes are 
redundant).
We recall that $d<\infty$ denotes the dimension of 
$\tilde \hr$ (see below \eref{39})
and $\tilde d_E\leqslant d$ the number of distinct energy 
eigenvalues $E_n$ of $\tilde H$ 
(see below \eref{4}).

Adopting the approach of Short and Farrelly \cite{sho12},
we start by considering the quantity 
$\langle \tilde \sigma^2(t)\rangle_T$,
where $\langle \cdot\rangle_T$ denotes a
temporal average over the time interval $[0, T]$ with
arbitrary but finite $T>0$. From 
\eref{6} and \eref{18} we can infer that
\begin{equation}
\bigl\langle \tilde \sigma^2(t)\bigr\rangle_T=
\Biggl\langle \Bigl|\sum_{m\not=n} \Tr\{\tilde\rho_{mn}\tilde A\}
\exp\left[-\rmi(E_m-E_n)t\right]\Bigr|^2 \Biggr\rangle_T,
\label{45}
\end{equation}
where $\tilde\rho_{mn}(0)$ is abbreviated as
$\tilde\rho_{mn}$ and the sum runs over the finite 
set of pairs of labels 
\begin{equation}
\gset:=\bigl\{(m,n)\, |\, m ,n\in[1,\dots,\tilde d_E],\, m\not=n\bigr\}.
\label{46}
\end{equation}
For any $\alpha=(m,n)\in\gset$ we define
\begin{equation}
G_\alpha := E_m-E_n,\qquad
v_\alpha := \Tr\{\tilde\rho_{mn}\tilde A\}.
\label{48}
\end{equation}
We thus can rewrite \eref{45} as
\begin{equation}
\bigl\langle \tilde \sigma^2(t)\bigr\rangle_T=
\Biggl\langle \Bigl|\sum_\alpha v_\alpha
\exp\left[-\rmi G_\alpha t\right]\Bigr|^2 \Biggr\rangle_T 
=\sum_{\alpha,\beta} v^\ast_\alpha M_{\alpha\beta} v_\beta,
\label{49}
\end{equation}
where we introduced the self-adjoint, non-negative,
finite-dimensional matrix 
$M$ with matrix elements
\begin{equation}
M_{\alpha\beta}:=\bigl\langle \exp\left[\rmi(G_\alpha-G_\beta)t\right]\bigr\rangle_T.
\label{50}
\end{equation}
Denoting by $\norm{M}$ the standard operator norm of the 
matrix $M$ (see \eref{10}), 
it follows that \cite{sho12}
\begin{eqnarray}
\bigl\langle \tilde \sigma^2(t)\bigr\rangle_T \leqslant S \norm{M}
\label{51}
\\
S := \sum_{\alpha} |v_\alpha|^2
=\sum_{m\not = n} \bigl|\Tr\{\tilde\rho_{mn}\tilde A\}\bigr|^2.
\label{52}
\end{eqnarray}

Bounds on the two factors $S$ and $\norm{M}$ in \eref{51} are
constructed in the following two subsections.

\subsection{Bound on $S$}
\label{s41}
We exploit \eref{2}, \eref{7}, and the cyclic
invariance of the trace to conclude that
\begin{eqnarray}
\Tr\{\tilde\rho_{mn}\tilde A\} & = &
\Tr\{\tilde P_m \tilde\rho \tilde P_n \tilde A \tilde P_m\}.
\label{53}
\end{eqnarray}
Similarly to the derivation in \eref{29}--\eref{31}, we
write
\begin{equation}
\tilde P_m \tilde\rho \tilde P_n \tilde A\tilde P_m = (\tilde P_m \tilde\rho^{1/2})(\tilde\rho^{1/2} \tilde P_n \tilde A\tilde P_m)
\end{equation}
from which it follows that
\begin{equation}
|\Tr\{\tilde\rho_{mn}\tilde A\}|^2
\leqslant \tilde p_m 
\Tr\{\tilde\rho_{nn} \tilde A \tilde P_m\tilde A\} 
.
\label{54}
\end{equation}

We first evaluate by means of \eref{54} 
all summands in \eref{52} with $n=1$,
\begin{eqnarray}
S_{n=1} & \leqslant & \sum_{m\geqslant 2}
\tilde p_m \Tr\{\tilde\rho_{11} \tilde A \tilde P_m\tilde A\}
\nonumber
\\
& \leqslant &
\max_{n\geqslant 2} p_n 
\Tr\Bigl\{\tilde\rho_{11} \tilde A \sum_{m\geqslant 2}\tilde P_m\tilde A\Bigr\}
\nonumber
\\
& \leqslant & 
\max_{n\geqslant 2}  p_n\Tr\{\tilde\rho_{11}\} \Bigl\|\tilde A \sum_{m\geqslant 2}\tilde P_m\tilde A\Bigr\|
\nonumber
\\
& \leqslant &
\max_{n\geqslant 2} p_n \norm{A}^2.
\label{55}
\end{eqnarray}
In the second line we used that 
$\tilde p_m\leqslant p_m \leqslant \max_{n\geqslant 2}p_n$ for all $m\geqslant 2$,
see \eref{43}.
The third line is based on a similar line of reasoning
as in \eref{33}, exploiting that $\tilde\rho_{11}$ is
a non-negative, self-adjoint operator.
In the last line we used that
$\Tr\{\tilde\rho_{11}\}=\tilde p_1\leqslant 1$
(see \eref{11} and \eref{41}),
that $\norm{BC}\leqslant \norm{B}  \norm{C}$
for arbitrary operators $B$, $C$ of finite norm, and that
$\sum_{m\geqslant 2}\tilde P_m$ 
is a projector and hence of unit norm.

For symmetry reasons, the same estimate as
in \eref{55} applies for the summands with
$m=1$ in \eref{52}.
The remaining summands in \eref{52} satisfy $m\not = n$
and $m, n\geqslant 2$. By including also those with $m=n$,
the sum can only increase, resulting in
\begin{equation}
S\leqslant 2 \norm{A}^2 {\max_n}' p_n +
{\sum_{m,n}}'  \bigl|\Tr\{\tilde\rho_{mn}\tilde A\}\bigr|^2,
\label{56}
\end{equation}
where the prime indicates that indices $1$ are excluded
from the maximization and the summation. 
Clearly, instead of this special index $1$ we could have 
selected any other index as well.
Thus, in agreement with \sref{s25},
the prime can and will be understood 
as excluding the index belonging to the 
maximally populated level.

The remaining sum in \eref{56} can be estimated
by means of the two non-negative, self-adjoint operators
\begin{equation}
\tilde\omega' := {\sum_n}'  \tilde \rho_{nn},\qquad
\tilde\omega'' := {\sum_m}'  \tilde p_m  \tilde P_m
\label{58}
\end{equation}
in the following way:
\begin{eqnarray}
{\sum_{m,n}}'  \bigl|\Tr\{\tilde\rho_{mn}\tilde A\}\bigr|^2 
\leqslant
\Tr\{\tilde\omega' \tilde A \tilde \omega'' \tilde A\} 
\leqslant
\sqrt{
\Tr\{(\tilde \omega')^2 \tilde A^2\} \Tr\{(\tilde \omega'')^2 \tilde A^2\}
}
\nonumber
\\
\leqslant
\sqrt{
\Tr\{(\tilde \omega')^2\} \norm{\tilde A}^2
\Tr\{(\tilde \omega'')^2\} \norm{\tilde A}^2
}
=
\norm{\tilde A}^2 \sqrt{{\sum_n}' \Tr\{\tilde \rho_{nn}^2\}
{\sum_m}' \tilde \mu_m \tilde p_m^2}.
\label{59}
\end{eqnarray}
In the first inequality, we exploited \eref{54}, the
next two ones follow by arguments similar
to the ones in \eref{31} and \eref{33}.
The last equation is based on
\eref{2}, \eref{5},
\eref{7} and \eref{58}.

Once again, the line of reasoning in \eref{59}
follows very closely that of Short and Farrelly 
\cite{sho11,sho12}.
The main difference is that these authors
focus, in a first step, solely on pure states
and only in the end extend their result 
to mixed states via purification.
Along this line, one actually arrives at a final result
which is slightly different from \eref{59}, namely
\begin{equation}
{\sum_{m,n}}'   |\Tr\{\tilde\rho_{mn}\tilde A\}|^2 
\leqslant \norm{\tilde A}^2 {\sum_n}' \tilde p_n^2.
\label{60}
\end{equation}
Closer inspection shows that \eref{59} and
\eref{60} agree if and only if all energies 
are non-degenerate.
In any other case, one can show that
the bound \eref{60} is sharper than \eref{59}.
On the other hand, one readily sees that
$\Tr\{\tilde\rho_{nn}^2\}\leqslant \tilde p_n^2$
and hence
\begin{equation}
\sqrt{{\sum_n}' \Tr\{\tilde \rho_{nn}^2\}
{\sum_m}' \tilde \mu_m \tilde p_m^2}
\leqslant
{\max_m}'\sqrt{\tilde \mu_m}\,{\sum_n}' \tilde p_n^2.
\label{61}
\end{equation}
The equality sign applies whenever $\rho(t)$ is a 
pure state contained in the maximally degenerate 
energy eigenspace,
in any other case the inequality sign applies.
In conclusion, \eref{60} outperforms \eref{59}
by at most a factor of ${\max_m'}\sqrt{\tilde \mu_m}$,
i.e.\ the square root of the maximal energy 
degeneracy.

The purification argument \cite{sho11,sho12}
behind \eref{60} is mathematically
very appealing (somewhat reminiscent of 
evaluating real integrals by `complexification')
but its physical content remains slightly
mysterious. 
It is reassuring that one can get at least
as far as \eref{59} without this argument,
but it is annoying that \eref{60} could not be 
fully recovered.

Working with the stronger bound \eref{60} 
by Short and Farrelly, the estimate \eref{56} for $S$
in combination with
\eref{14} and \eref{43}
takes on the simple form
\begin{equation}
S\leqslant 3 \norm{A}^2 {\max_n}' p_n.
\label{62}
\end{equation}
While the maximization so far includes only
a finite number of indices $n$,
this restriction can be readily released,
as the maximum can only increase in this way.
As in \sref{s25}, the last factor then 
represents the second largest level population
of the original, possibly infinite-dimensional system.

\subsection{Bound on $\norm{M}$ in terms of energy gaps}
\label{s42}
The main idea is that, since $M$ is a finite-dimensional
matrix, its operator norm $\norm{M}$ converges 
towards a well-defined limit as the averaging
time $T$ (see \eref{45}, \eref{50}) tends to
infinity.
Hence, $\norm{M}$ can be readily bounded from above 
for all sufficiently large (but finite) $T$.

Quantitatively, by
setting out from the inequality
\begin{equation}
\norm{M}\leqslant \max_\beta\sum_\alpha |M_{\alpha \beta}|,
\label{63}
\end{equation}
Short and Farrelly \cite{sho12} derived the estimate
\begin{equation}
\norm{M}\leqslant N(\epsilon)
\left(1+\frac{8\log_2\! \tilde d_E}{\epsilon T}\right)
\label{64}
\end{equation}
for arbitrary $\epsilon>0$
and $T>0$.
Here, $\tilde d_E$ is the number of distinct energy 
eigenvalues $E_n$
(see below \eref{44}) and 
\begin{equation}
N(\epsilon) :=
\max_{E} \bigl|\{ \alpha \in \gset \, | \, G_\alpha \in[E,E+\epsilon)\}\bigr|.
\label{65}
\end{equation}
According to \eref{46} and \eref{48}, $N(\epsilon)$ is 
thus the maximum number of energy gaps
$G_\alpha=E_m-E_n$ in any interval of size 
$\epsilon$.

The main implication of this result can also be deduced 
directly from \eref{63}: since we are dealing
with finite-dimensional systems, the 
{\em finite number} of all matrix
elements $|M_{\alpha\beta}|$ in \eref{50} 
with $G_\alpha\not=G_\beta$ can be 
{\em simultaneously} bounded by an arbitrarily 
small upper limit for sufficiently large $T$.
Hence, their contribution to \eref{63}
can be made smaller than the contribution
of all the remaining summands, satisfying
$G_\alpha=G_\beta$ and thus $M_{\alpha\beta}=1$. 
It follows for all sufficiently large $T$ that
\begin{equation}
\norm{M}\leqslant 2g,
\label{66}
\end{equation}
where
\begin{equation}
g := 
\max_\beta\left|\{ \alpha\in \gset \, | \, G_\alpha=G_\beta\}\right|
\label{67}
\end{equation}
denotes the maximal degeneracy of
energy gaps. Note that
only the energy eigenvalues $E_n$ of the
restricted, finite-dimensional Hamiltonian 
$\tilde H$ contribute to $\gset$ in \eref{46}
and hence to the degenerate energy gaps counted
in \eref{67}.

\subsection{Problems in the infinite-dimensional limit}
\label{s43}
So far, we have assumed a finite dimensionality 
$d$ of the Hilbert space $\tilde \hr$.
In this subsection we argue that it is intuitively 
suggestive that everything `should go well'
upon letting $d$ go to infinity, but that a more
rigorous justification is problematic. From 
the latter viewpoint, the considerations
in \sref{s3} are thus indispensable for
infinite-dimensional systems.

First of all, since $A$ has a finite range $\da$,
see \eref{8}, it follows that $\sigma(t)$
from \eref{17} is contained in the finite interval
$[-2\da, 2\da]$ for all times $t$.
This suggest (but does not prove) that the temporal
average $\langle \sigma^2(t)\rangle_T$ converges
in the limit $T\to\infty$ even for infinite 
dimensional systems:
although one can readily construct mathematical
examples of bounded functions without
a well-defined infinite-time average,
it appears plausible that `reasonable'
physical models will result in functions $\sigma^2(t)$
which do not exhibit the pathologies of those 
examples.

On the other hand, for any given finite dimension $d$,
equation \eref{49} has a well defined 
$T\to\infty$ limit.
Focusing on the simplest case of non-degenerate energy
gaps, $M_{\alpha\beta}$ in \eref{50}
approaches $\delta_{\alpha\beta}$ and hence the
inequality \eref{51} turns into an equality 
with $\norm{M}=1$.
Since $S$ in \eref{52} is positive and
bounded by the $d$-independent estimate \eref{62},
it is once again suggestive that $S$ itself converges
for $d\to\infty$.
Under the further assumption that the two limits
$T\to\infty$ and $d\to\infty$ commute,
one then readily finds an upper bound analogous to
\eref{68} for infinite dimensions and
all sufficiently large, but finite, $T$.

Although these heuristic arguments appear plausible
at first glance, some subtle open questions
remain upon closer inspection:
For infinite-dimensional systems
one typically expects the existence
of arbitrarily small, but non-vanishing
energy gap differences $G_\alpha-G_\beta$.
While each single matrix element \eref{50}
then still converges for $T\to\infty$,
the same is no longer clear for the entire,
infinite-dimensional
matrix $M$ and/or its norm $\norm{M}$.
For the same reason, $N(\epsilon)$ from \eref{65}
is expected to diverge for $d\to\infty$
and any fixed $\epsilon>0$, so that
\eref{64} becomes useless.
Likewise, our derivation of \eref{66}
breaks down.
In other words, interchanging the limits
$T\to\infty$ and $d\to\infty$ is
a rather delicate, unsettled issue.

\section{Main result and discussion}
\label{s5}
Combining \eref{19}, 
\eref{51}, 
\eref{62} and \eref{66},
we obtain the inequality 
\begin{equation}
\bigl\langle \tilde \sigma^2(t)\bigr\rangle_T
\leqslant  \case{3}{2} g\da^2 {\max_n}' p_n 
\label{68}
\end{equation}
for all sufficiently large $T$.

For any given $\epsilon>0$ and $T>0$
we define the measure 
of all times $t\in [0, T]$ for which $|\tilde\sigma(t)|\geqslant\epsilon$ holds true,
\begin{equation}
\tilde T_\epsilon := 
\lambda\big( \{t\, | \, t\in [0,T] \ \mbox{and}\ 
|\tilde\sigma(t)|\geqslant\epsilon\}\big),
\label{69}
\end{equation}
where $\lambda$ denotes the Lebesgue measure.
It follows that $\tilde\sigma^2(t)\geqslant \epsilon^2$
for a set of times $t$ of measure $\tilde T_\epsilon$
and
$\tilde\sigma^2(t)\geqslant 0$ for all remaining times $t$
in $[0, T]$.
Hence the temporal average of 
$\tilde\sigma^2(t)$
over the time interval $[0, T]$ must be at least
$\epsilon^2 \tilde T_\epsilon/T$,
\begin{equation}
\bigl\langle \tilde\sigma^2(t)\bigr\rangle_T\geqslant 
\frac{\epsilon^2\tilde T_\epsilon}{T}.
\label{70}
\end{equation}
Choosing $\epsilon=\dda/2$, we can conclude from \eref{68}
and \eref{70} that
\begin{equation}
\frac{\tilde T_{\dda/2}}{T}\leqslant 
6g\left(\frac{\da}{\dda}\right)^2 {\max_n}' p_n 
\label{71}
\end{equation}
for all sufficiently large $T$.

Next we infer from 
\eref{36}, \eref{39} and \eref{44}
that
\begin{equation}
|\Tr\{\rho(t)A\}-\Tr\{\omega A\}|\leqslant |\tilde\sigma(t)|+\frac{\dda}{2}.
\label{72}
\end{equation}
Analogously to \eref{69}, we define the measure
of all times $t\in [0, T]$ with the
property that $|\Tr\{\rho(t)A\}-\Tr\{\omega A\}|$
exceeds the experimental resolution limit $\dda$,
\begin{equation}
T_{\dda} := 
\lambda \big( \{t\, | \, t\in [0,T] \ \mbox{and}\ 
|\Tr\{\rho(t)A\}-\Tr\{\omega A\}|\geqslant\dda\}\big).
\label{73}
\end{equation}
In view of \eref{69} with $\epsilon=\dda/2$ and \eref{72}, we conclude that 
$T_{\dda}\leqslant\tilde T_{\dda/2}$. With \eref{71} we thus 
arrive at the main result of our present work,
\begin{equation}
\frac{T_{\dda}}{T}\leqslant 
6 g\left(\frac{\da}{\dda}\right)^2 {\max_n}' p_n 
\label{74}
\end{equation}
for all sufficiently large $T$.

The left hand side of \eref{74}
represents the fraction of all times
$t\in[0, T]$ for which there is an experimentally
resolvable difference between the true expectation 
value $\Tr\{\rho(t)A\}$ and the time-independent 
`equilibrium expectation value' $\Tr\{\omega A\}$.

On the right hand side, $g$ is the maximal degeneracy 
of energy gaps from \eref{67}, i.e.\ the
maximal number of (exactly) coinciding energy 
differences among all possible pairs of 
distinct energy eigenvalues
of the reduced, finite-dimensional Hamiltonian 
$\tilde H$ from \sref{s3}, and
as such is determined by properties of both the
Hamiltonian $H$ and the initial condition $\rho(0)$.
Alternatively, one may also take into account
all energies of the full system Hamiltonian
$H$, since the maximum can only increase in 
this way, but this increase might possibly
become prohibitively huge for infinite 
dimensional systems (see also \sref{s6}).

The next factor $\da/\dda$ appearing in \eref{74}
is the range-to-resolution ratio
of $A$ from \sref{s23},
i.e.\ a characteristic property of the observable 
$A$ only, and can be considered as bounded according 
to \eref{9} for all experimentally realistic 
measurements $A$.
Going back to \sref{s41}, one
readily sees that one could as well 
replace $\da$ by the range of the reduced 
observable $\tilde A$. This range typically is
somewhat smaller than the original range $\da$ of $A$ (see \eref{43}),
but this gain might often not 
be worth the effort.

Finally, ${\max_n}' p_n$ in \eref{74} stands for 
the second-largest, ensemble-averaged oc\-cu\-pa\-tion 
probability of the (possibly degenerate) energy 
eigenvalues $E_n$, see sections \ref{s24} and \ref{s25}.
Similarly as for $A$, one alternatively could maximize
over the reduced level populations $\tilde p_n$, but
often this will not be worthwhile.
Essentially, ${\max_n}' p_n$ is thus a
characteristic property of the initial 
condition $\rho(0)$,
but obviously also the Hamiltonian $H$ itself
matters.
Typically, one expects that the rough upper
bound \eref{13} applies, except if certain
energy eigenvalues
are so extremely highly degenerate that
the multiplicities defined in \eref{5} severely
reduce the pertinent energy level density
compared to the non-degenerate case, 
see \sref{s24}.

For a system with sufficiently many degrees of freedom
$f$ and no exceedingly large degeneracy
of the energy eigenvalues and the energy gaps, 
we thus can conclude from \eref{74} with 
\eref{9} and \eref{13} 
that the system behaves in every possible experiment 
exactly as if it were in the equilibrium
state $\omega$ for the overwhelming majority of times
within any sufficiently large (but finite) 
time interval $[0, T]$.
In particular, $T$ must obviously be much larger
than the relaxation time in case
of a far-from-equilibrium initial 
condition $\rho(0)$.
A more detailed quantitative bound on $T$ 
follows from the result \eref{64} by 
Short and Farrelly \cite{sho12}.

We emphasize that even in the absence of any
measurable difference between $\rho(t)$ and
$\omega$, the equilibrium state $\omega$
itself is never realized in the actual 
system, and as such is a purely formal,
theoretical construct.
In particular, the difference between 
$\Tr\{\rho(t)A\}$ and $\Tr\{\omega A\}$
is not a quantity one can measure in a real 
system, not even in principle.

Further interesting physical implications of 
\eref{74} are discussed, in e.g., \cite{rei10}.

\section{Conclusions}
\label{s6}
To summarize, by adopting and extending recent
works by Short \cite{sho11} and by Short and 
Farrelly \cite{sho12}, we demonstrated
equilibration of isolated macroscopic quantum 
systems in the sense 
that deviations from a time-independent
steady state are unmeasurably 
small for the overwhelming majority of
times within any sufficiently large
time interval.
This conclusion applies for arbitrary
systems with (countably) infinite 
dimensions,
initial states exhibiting a macroscopic 
population of at most one energy level,
and Hamiltonians without exceedingly
large degeneracies of energy eigenvalues
and energy gaps.

As soon as a model includes at least one
continuous degree of freedom (e.g.\ a spatial
coordinate), the pertinent Hilbert space
is necessarily of infinite dimension.
If the system is furthermore of finite spatial 
extension (e.g.\ due to confining walls),
the Hamiltonian $H$ features a discrete 
spectrum and can be written in 
the form \eref{1}.
Both conditions are clearly satisfied 
in almost any model which is
not based on some extreme 
simplifications.
Moreover, it is practically impossible
to prepare a real system so that only
a finite number of 
energy eigenstates is populated.
Rather, generically an infinite number of them
contributes with
possibly small, yet finite, amplitudes.
Reducing or truncating this situation
in whatever way to finite dimensions
is especially problematic 
with respect to rigorous 
statements about the exact quantum 
mechanical evolution over arbitrarily 
long time intervals.
More precisely, interchanging the
limit of infinite dimensions with
the long time limit, as discussed in 
\sref{s43}, is a subtle problem, 
justifying the detailed treatment of the 
infinite-dimensional case in \sref{s3}.

Admitting systems with degenerate energy eigenvalues
and energy gaps are among the most important
steps forward achieved in \cite{sho11,sho12}.
On the other hand, it is generally taken for
granted that such degeneracies are absent in 
generic Hamiltonians, see e.g.\
\cite{per84,sre99,tas98,rei08,lin09,rei10}
and, in particular, section 3.2.1 of 
\cite{gol06} and references therein.
Roughly speaking, Hamiltonians with such degeneracies
are of measure zero compared to `all' Hamiltonians.
They only arise in the presence of special reasons 
like (perfect) symmetries, additional conserved 
quantities, or fine-tuning of parameters,
which can be ruled out for typical real 
systems provided they cannot be further decomposed 
into non-interacting subsystems \cite{lin09}.
It is reassuring that even some of those exceptions---namely
those without exceedingly high degeneracies---are 
now covered in \cite{sho11,sho12} 
and in the present work. 
Accidental degeneracies due to fine-tuning of parameters
should not lead to high degeneracies, but the quantitative 
effect of symmetries on level and gap degeneracies is not 
clear to the present authors.
In order to better understand the occurrence of degeneracies, 
it appears indispensable to study specific examples. 
The harmonic oscillator would be one which 
is still not admissible (as the degeneracy of energy 
gaps is too large), but the hope that other relevant 
examples will be `tame' enough seems reasonable.

\ack
\addcontentsline{toc}{section}{Acknowledgments}
PR acknowledges financial support by the
Deutsche Forschungsgemeinschaft 
under grant RE1344/7-1. MK acknowledges financial 
support by the Incentive Funding for Rated 
Researchers programme of the National Research 
Foundation of South Africa.


\bibliographystyle{unsrtnat}

\begin{thebibliography}{41}
\addcontentsline{toc}{section}{References}

\bibitem{skl93}
Sklar L 1993 {\em Physics and Chance} (Cambridge: Cambridge University Press)

\bibitem{rei08}
Reimann P 2008 \PRL {\bf 101} 190403

\bibitem{lin09}
Linden N, Popescu S, Short A J and Winter A 2009 \PR E {\bf 79} 061103

\bibitem{lin10}
Linden N, Popescu S, Short A J and Winter A 2010 \NJP {\bf 12} 055021

\bibitem{rei10}
Reimann P 2010 \NJP {\bf 12} 055027

\bibitem{sho11}
Short A J 2011 \NJP {\bf 13} 053009

\bibitem{sho12}
Short A J and Farrelly T C 2012 \NJP {\bf 14} 013063

\bibitem{gol10a}
Goldstein S, Lebowitz J L, Mastrodonato C, 
Tumulka R and Zangh\`{\i} N 2010 \PR E {\bf 81} 011109

\bibitem{gol10b}
Goldstein S, Lebowitz J L, Tumulka R and Zangh\`{\i} 2010
\EJP H {\bf 35} 173

\bibitem{gol10c}
Goldstein S, Lebowitz J L, Mastrodonato C, 
Tumulka R and Zangh\`{\i} N 2010
{\em Proc. R. Soc. London Ser. A} {\bf 466} 3203

\bibitem{tas10}
Tasaki H 2010 The approach to thermal equilibrium and 
`thermodynamic nor\-mal\-i\-ty' arXiv:1003.5424v4

\bibitem{tas98}
Tasaki H 1998 \PRL {\bf 80} 1373

\bibitem{sre99}
Srednicki M 1999 \JPA {\bf 32} 1163

\bibitem{caz06}
Cazalilla M A 2006 \PRL {\bf 97} 156403

\bibitem{rig07}
Rigol M, Dunjko V, Yurovsky V and Olshanii M 2007
\PRL {\bf 98} 050405

\bibitem{col07}
Kollath C, L\"auchli A M and Altman E 2007 
\PRL {\bf 98} 180601

\bibitem{man07}
Manmana S R, Wessel S, Noack R M and Muramatsu A 2007 
\PRL {\bf 98} 210405

\bibitem{cra08}
Cramer M, Dawson C M, Eisert J and Osborne T J 2008
\PRL {\bf 100} 030602

\bibitem{bar08}
Barthel T and Schollw\"ock U 2008 
\PRL {\bf 100} 100601

\bibitem{rig08}
Rigol M, Dunjko V and Olshanii M 2008 
{\em Nature} {\bf 452} 854

\bibitem{spec}
Focus Issue on {\em Dynamics and thermalization in isolated 
quantum many-body systems},
edited by Cazalilla M A and Rigol M 2010
\NJP {\bf 12}

\bibitem{pal10}
Pal A and Huse D A 2010 \PR B {\bf 82} 174411

\bibitem{ban11}
Ba\~{n}uls M C, Cirac J I and Hastings M B 2011
\PRL {\bf 106} 050405

\bibitem{pol11}
Polkovnikov A, Sengupta K, Silva A and Vengalattore M 2011
\RMP {\bf 83} 863

\bibitem{camp11}
Campos Venuti L, Jacobson N T, Santra S and Zanardi P 2011
\PRL {\bf 107} 010403

\bibitem{jacobson11}
Jacobson N T, Campos Venuti L and Zanardi P 2011
\PR A {\bf 84} 022115

\bibitem{kas11}
Kastner M 2011 \PRL {\bf 106} 130601

\bibitem{kas12}
Kastner M 2012 {\em Cent. Eur. J. Phys.} DOI:10.2478/s11534-011-0122-4 (in press)

\bibitem{gog11}
Gogolin C, M\"uller M P and Eisert J 2011
\PRL {\bf 106} 040401

\bibitem{ike11}
Ikeda T N, Watanabe Y and Ueda M 2011
\PR E {\bf 84} 021130

\bibitem{realobs1}
Khinchin A Y 1960 {\em Mathematical Foundations of Quantum Statistics}
(New York: Graylock Press) chapter 3

\bibitem{realobs2}
Wigner E P 1963 {\em Am. J. Phys.} {\bf 31} 6

\bibitem{realobs3}
van Kampen N G 1992 {\em Stochastic Processes in Physics and Chemistry}
(Amsterdam: Elsevier) chapter XVII.7

\bibitem{realobs4}
Sugita A 2007 {\em Nonlinear Phenom. Complex Syst.} {\bf 10} 192

\bibitem{realobs5}
Penrose O 1970 {\em Foundations of Statistical Mechanics} (Oxford:
Pergamon) chapter 1

\bibitem{lof}
Martin-L\"of A 1979 {\em Statistical Mechanics and the Foundations of Thermodynamics
(Lecture Notes in Physics vol 101)} (Berlin: Springer)

\bibitem{geo95}
Georgii H-O 1995 {\em J. Stat. Phys.} {\bf 80} 1341

\bibitem{pop06} 
Popescu S, Short A J and Winter A 2006 {\em Nature Phys.} {\bf 2} 754

\bibitem{rei08a}
Reimann P 2008 {\em J. Stat. Phys.} {\bf 132} 921

\bibitem{lldiu}
Diu B, Guthmann C, Lederer D and Roulet B 1996 
{\em Elements de Physique Statistique} (Paris: Hermann)

\bibitem{hob71}
Hobson A 1971 {\em Concepts in Statistical Mechanics}
(New York: Gordon and Breach)

\bibitem{per84}
Peres A 1984 \PR A {\bf 30} 504

\bibitem{gol06}
Goldstein S, Lebowitz J L, Tumulka R and Zangh\`{\i} N 2006
{\em J. Stat. Phys.} {\bf 125} 1197

\end{thebibliography}

\end{document}